\documentclass[sigconf]{acmart}
\usepackage{extarrows}
\usepackage{color} 
\usepackage{xcolor}
\usepackage{makecell}
\usepackage{amsmath,amsfonts}
\usepackage{algorithmic}
\usepackage{array}
\usepackage{graphicx}
\usepackage{stfloats}
\usepackage{multicol}
\usepackage{multirow}
\usepackage{hhline}
\usepackage{colortbl}
\usepackage{bm}
\usepackage{url}
\usepackage{enumitem}
\definecolor{customgreen}{rgb}{0.3647, 0.6784, 0.3294}

\AtBeginDocument{%
  \providecommand\BibTeX{{%
    \normalfont B\kern-0.5em{\scshape i\kern-0.25em b}\kern-0.8em\TeX}}}

\copyrightyear{2023} 
\acmYear{2023} 
\setcopyright{acmlicensed}\acmConference[MM '23]{Proceedings of the 31st ACM International Conference on Multimedia}{October 29-November 3, 2023}{Ottawa, ON, Canada}
\acmBooktitle{Proceedings of the 31st ACM International Conference on Multimedia (MM '23), October 29-November 3, 2023, Ottawa, ON, Canada}
\acmPrice{15.00}
\acmDOI{10.1145/3581783.3611817}
\acmISBN{979-8-4007-0108-5/23/10}

\begin{document}
\begin{sloppypar}
\title{Target-Guided Composed Image Retrieval}

\author{Haokun Wen}
\affiliation{
  \institution{\normalsize Harbin Institute of Technology (Shenzhen)}
  \city{Shenzhen}
  \country{China}
  }
\email{whenhaokun@gmail.com}

\author{Xian Zhang}
\affiliation{%
  \institution{\normalsize Harbin Institute of Technology (Shenzhen)}
  \city{Shenzhen}
  \country{China}
  }
\email{zhangxianhit@gmail.com}

\author{Xuemeng Song}
\authornote{Corresponding authors: Xuemeng Song and Liqiang Nie.}
\affiliation{%
  \institution{\normalsize Shandong University}
  \city{Qingdao}
  \country{China}
  }
\email{sxmustc@gmail.com}

\author{Yinwei Wei}
\affiliation{%
  \institution{\normalsize Monash University}
  \city{Melbourne}
  \country{Australia}
  }
\email{weiyinwei@hotmail.com}

\author{Liqiang Nie}
\authornotemark[1]
\affiliation{%
  \institution{\normalsize Harbin Institute of Technology (Shenzhen)}
  \city{Shenzhen}
  \country{China}
  }
\email{nieliqiang@gmail.com}





\begin{abstract}

Composed image retrieval (CIR) is a new and flexible image retrieval paradigm, which can retrieve the target image for a multimodal query, including a reference image and its corresponding modification text. 
Although existing efforts have achieved compelling success, they overlook the conflict relationship modeling between the reference image and the modification text for improving the multimodal query composition and the adaptive matching degree modeling for promoting the ranking of the candidate images that could present different levels of matching degrees with the given query. 
To address these two limitations, in this work, we propose a Target-Guided Composed Image Retrieval network (TG-CIR). In particular, TG-CIR first extracts the unified global and local attribute features for the reference/target image and the modification text with the contrastive language-image pre-training model (CLIP) as the backbone, where an orthogonal regularization is introduced to promote the independence among the attribute features. Then TG-CIR designs a target-query relationship-guided multimodal query composition module, comprising a target-free student composition branch and a target-based teacher composition branch, where the target-query relationship is injected into the teacher branch for guiding the conflict relationship modeling of the student branch. Last, apart from the conventional batch-based classification loss, TG-CIR additionally introduces a batch-based target
similarity-guided matching degree regularization to promote the metric learning process. Extensive experiments on three benchmark datasets demonstrate the superiority of our proposed method.
  
\end{abstract}



\maketitle

\section{Introduction}
\begin{figure}[t]
	\includegraphics[width=0.82\linewidth]{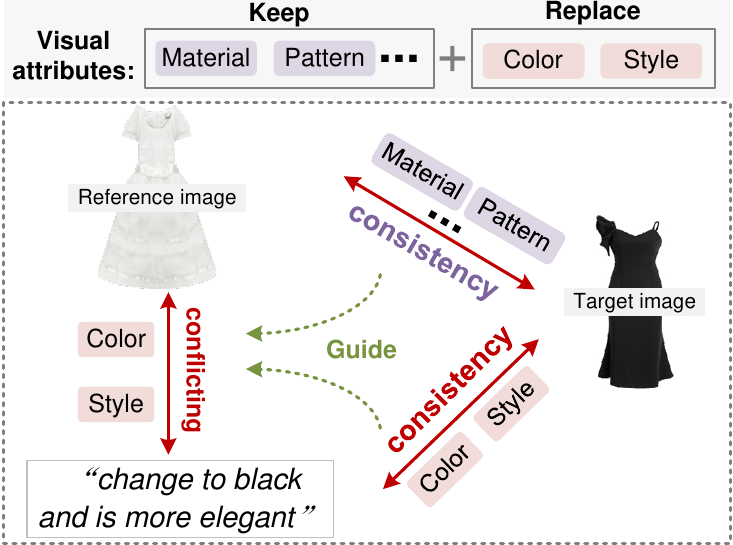}
    \vspace{-0.5em}
	\caption{Relationships among the reference image, modification text, and target image in CIR. }
    \vspace{-1.2em}
	\label{fig:intro}
\end{figure}


Different from the conventional image retrieval paradigm that only supports either the pure text query~\cite{textbased1,raoj}  or the pure image query~\cite{imagebased1,imagebased2}, composed image retrieval (CIR)~\cite{tirg} allows users to input the multimodal query, including a reference image and its corresponding modification text, to express their search intentions more flexibly. Essentially, the reference image can basically but cannot exactly capture the user's search demand, where a modification text is needed to describe the difference between the reference image and the user's desired one in his/her mind. Given its great potential value in various real-world applications, such as commercial product search~\cite{wei2,liuh} and interactive intelligent robots~\cite{xl1,robot2,zhangx}, CIR has garnered increasing research attention in recent years.

Essentially, the key to an effect CIR system lies in two key components: (1) multimodal query composition for accurately capturing the user's search demands, and (2) metric learning for accurately ranking the candidate images. 

For multimodal query composition, existing methods devote to design various neural networks~\cite{tirg,val,cosmo,hffca,artemis,clvcnet,clip4cir,guanw,guan2,xl3} to compose the multimodal query, but overlook to model the intrinsic conflicting relationship between the multimodal query. Figure~\ref{fig:intro} illustrates an example of multimodal query, where the reference image indicates that the user may want a white princess dress, while the modification text specifies that the user wants to change the color and style of the reference image to ``black'' and ``more elegant'', respectively. In this case, the reference image and the modification text conflict with regard to the visual attributes of color and style. Apparently, conducting the conflicting relationship modeling contributes to learn which  visual attributes of the reference image should be kept as well as which should be replaced with those specified in the modification text, and thus improves the multimodal query composition.



\begin{figure}[t]
	\includegraphics[width=0.9\linewidth]{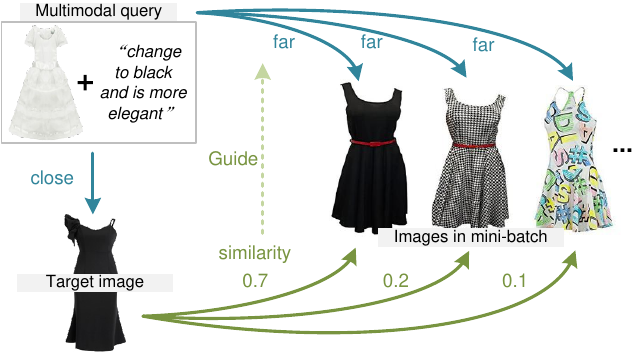}
    \vspace{-0.5em}
	\caption{Metric learning for CIR. The blue lines denote traditional metric learning, and the green lines refer to target similarity distribution-guided metric learning.}
    \vspace{-1.2em}
	\label{fig:intro2}
\end{figure}

As for metric learning, existing methods mainly use the triplet training data (\textit{i.e.}, \textless reference image, modification text, target image\textgreater) and adopt the batch-based classification loss function~\cite{tirg,qint} for optimization. In this manner, only the ground-truth target image is treated as the positive sample, while all other candidate target images in the mini-batch are treated as negative samples. Accordingly, only the ground-truth target image is forced to be the closest match to the multimodal query within a batch,
as illustrated by the blue lines in Figure~\ref{fig:intro2}. Nevertheless, this manner neglects that the other candidate target images may present different levels of matching degrees with the multimodal query. For instance, the first image (black dress) in the mini-batch matches the multimodal query better than the third image (colorful dress).
Moreover, it has been proven that due to the expensive cost for CIR training triplets annotation, the public dataset could contain false-negative samples for a given multimodal query~\cite{cirr}, namely the candidate images that match the multimodal query but are not annotated. Therefore, directly treating all the other candidate images as negative samples can adversely affect the metric learning process.


To address the limitations mentioned above, we aim to incorporate the conflict relationship modeling in the multimodal query composition and the adaptive matching degree modeling in the metric learning to promote the CIR performance.
Nevertheless, this is non-trivial due to the following three challenges.
1) Intuitively, to pave the way for the conflicting relationship modeling for multimodal query composition, we first need to unify the attribute features of the three elements (\textit{i.e.}, reference image, modification text, and target image) in CIR. Nevertheless, we do not have the explicit supervision for attribute feature extraction. Therefore, how to extract the unified attribute features for the three elements  without explicit supervision forms the first challenge.
2) The conflicting relationship between the multimodal query can be also manifested through the target-query relationships, \textit{i.e.},  ``reference image $\xleftrightarrow{\tiny{consistency}}$ target image'' and ``modification text $\xleftrightarrow{\tiny{consistency}}$ target image''. 
As shown in Figure~\ref{fig:intro}, by comparing the consistency between the target image and the modification text, we can learn that both color and style are the to-be-replaced attributes, while the other attributes should be kept by referring the consistency between the target image and the reference image. 
Nevertheless, the target image is unavailable during the testing stage. Therefore, how to feasibly utilize the target-query relationship to boost the multimodal query composition effect is another challenge. And 3) as illustrated by the green lines in Figure~\ref{fig:intro2}, the visual similarity between the ground-truth target image and the other candidate images in the mini-batch can provide guidance for the metric learning process. Intuitively, the more similar the candidate image is to the given ground-truth target image, the higher matching degree should be also assigned to it for the given multimodal query. Accordingly, how to effectively leverage the visual similarity between the ground-truth target image and the other candidate images in the mini-batch to achieve adaptive matching degree modeling is a crucial challenge.

To tackle these challenges, we present a Target-Guided Composed Image Retrieval network (TG-CIR), which comprises three key modules: attribute feature extraction, target-query relationship-guided multimodal query composition, and target similarity distribution-guided metric learning. The first module is to extract the unified attribute features of the reference/target image and modification text from both local and global perspectives, where an orthogonal regularization is designed to guarantee the attribute feature independence without explicit supervision signals for attribute feature extraction. The second module focuses on composing the attribute features of the multimodal query, where the conflicting relationship is modeled via the ``\textit{keep-and-replace}'' strategy. In particular, we design this module with two branches: target-based teacher composition branch and target-free student composition branch, where the target-query relationship is directly adopted as the guidance for the conflicting relationship modeling in the teacher branch, while the student branch is forced to mimic the teacher branch to gain the guidance indirectly. 
The third module aims to regularize the conventional \mbox{batch-based} classification loss function through a newly proposed \mbox{batch-based} target \mbox{similarity-guided} matching degree regularization. Extensive experiments on three real-world datasets validate the superiority of our model. 

The contribution of this work can be summarized in three-folds:
 \begin{itemize}[leftmargin=18pt]
     \item We propose a target-query relationship-guided multimodal query composition module with the ``\textit{keep-and-replace}'' paradigm. To the best of our knowledge, we are the first to leverage the target-query relationship to more effectively model the conflicting relationship in the multimodal query. 
     \item We propose a batch-based target similarity-guided matching degree regularization that can improve the performance of metric learning for CIR. As far as we know, we are the first to leverage the target visual similarity to promote the matching degree learning. 
     \item We propose an attribute feature extraction module, which can extract unified attribute features of the three elements of the CIR task from both local and global perspectives, to facilitate the conflicting relationship modeling. We have released our codes to facilitate other researchers\footnote{\url{https://anosite.wixsite.com/}tg-cir.}.

\end{itemize}


\section{Related Work}
Our work is closely in line with CIR and knowledge distillation.

\subsection{Composed Image Retrieval}
According to the type of feature extraction backbone, existing CIR methods can be generally classified into two groups. The first group utilizes traditional models with limited parameters, and the extracted features of the image and the text are usually in two separate modality spaces~\cite{tirg,val,clvcnet,amc,cosmo,datir,wei1,xl2}. A typical example is TIRG\cite{tirg}, which utilizes ResNet\cite{resnet} and LSTM\cite{lstm} to extract the image feature and the text feature, respectively. The designed gating and residual module are then used to compose the multimodal query features, where the reference image is adaptively preserved and transformed to derive the composed query representation. The batch-based classification loss is utilized to optimize the metric learning process. Following that, Chen \textit{et al.}~\cite{val} resorted to the attention mechanism
to fuse the hierarchical reference image features from the middle layers of ResNet with the
modification text feature. 
In contrast, the second group utilizes the multimodal pre-trained large model, like CLIP, as the feature extraction backbone, which has become mainstream recently~\cite{clip4cir, fame_vil}. These methods usually yield better performance due to the outstanding feature extraction ability on multimodal data. For example, Baldrati \textit{et al.}~\cite{clip4cir} resorted to CLIP to extract the image and text features, and then utilized a simple combiner module to fuse the multimodal query features to achieve a satisfactory retrieval performance. Recently, Han \textit{et al.}~\cite{fame_vil} designed a unified visual-linguistic model to handle multiple multimodal learning tasks in the fashion domain. Different cross-attention adaptors are utilized to adapt the model to different tasks, including CIR. Benefiting from the large model and multi-task learning, it has achieved a leading CIR performance. 

Although these methods have made prominent progress, they only rely on the multimodal query to conduct the multimodal query composition. Additionally, the commonly used batch-based classification loss cannot effectively rank the candidate images. Beyond these studies, our TG-CIR can leverage the visual information of the target image to guide both multimodal query composition and metric learning. 

\subsection{Knowledge Distillation}
Knowledge distillation~\cite{kl_hinton} is to transfer knowledge from a high-performing teacher model to a weak student network. The knowledge can be either the features from the middle layers~\cite{hint} or the logits from the output layer~\cite{kl_hinton,rao2}. This simple yet effective technique has drawn many researchers' attention. For example, Zhang \textit{et al.}~\cite{kl_zhang} utilized the logits of a pre-trained teacher model to guide the learning of the student model in the person search task. 
In addition, Wang \textit{et al.}~\cite{kl_wang} employed the features from the previous model as knowledge to guide the learning of another model on a new task in a continuous learning paradigm, which can achieve better performance on the new task. In this paper, considering that the target image is not available in the testing stage, we design a target-free student composition branch and a target-based teacher composition branch. We then resort to the knowledge distillation technique in the training stage to transfer knowledge from the teacher branch to the student branch. Finally, the student branch will be used in practice.

\section{Methdology}
In this section, we first formulate the research problem and subsequently elaborate on the proposed TG-CIR.

\begin{figure*}[htp]
	\includegraphics[width=0.93\linewidth]{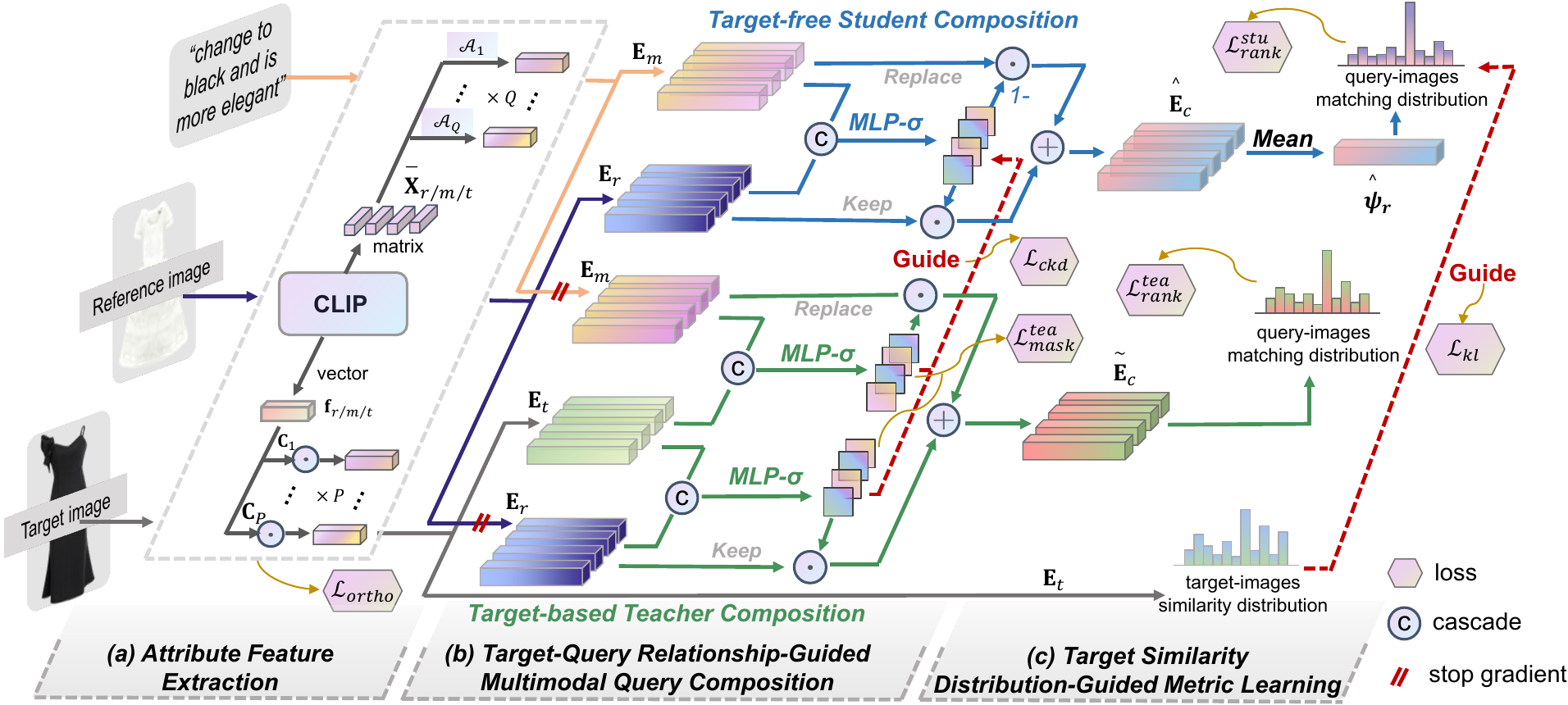}
    \vspace{-0.5em}
	\caption{The proposed TG-CIR consists of three key modules: (a) attribute feature extraction, (b) target-query relationship-guided multimodal query composition, and (c) target similarity distribution-guided metric learning.}
    \vspace{-1em}
	\label{fig:overall}
\end{figure*}

\subsection{Problem Formulation}
In this work, we aim to address the challenging CIR task, which can be formally defined as given a multimodal query consisting of a reference image and its modification text, the goal is to retrieve the corresponding target image from a set of gallery images.
Suppose we have a set of triplets, denoted as $\mathcal{D}=\left\{\left(x_{r},t_{m}, x_{t}\right)_{i}\right\}_{i=1}^{N}$, where $x_{r}$, $t_{m}$, and $x_{t}$ refer to the reference image, the modification text, and the target image, respectively. $N$ is the total number of triplets.
Based on $\mathcal{D}$, 
we aim to optimize a multimodal query composition function that can map the multimodal query $\left(x_r, t_m\right)$ into a latent metric space, in which its distance to its corresponding target image should be close.
This can be formally defined as follows,
\begin{equation}
    \mathcal{F}\left(x_r, t_m\right) \rightarrow \mathcal{H}\left(x_t\right),\label{eq1}
\end{equation}
where $\mathcal{F}$ represents the multimodal feature composition function of the multimodal query, while $\mathcal{H}$ denotes the feature embedding function for the target image.

\subsection{TG-CIR}
As the major novelty, we propose to exploit the visual information of the target image to promote the two essential components of composed image retrieval, \textit{i.e.}, the multimodal query composition and the metric learning. Specifically, we propose a Target-Guided Composed Image Retrieval network (TG-CIR), as illustrated in Figure~\ref{fig:overall}.
It consists of three key modules: (a) attribute feature extraction, (b) target-query relationship-guided multimodal query composition, and (c) target similarity distribution-guided metric learning. The first module works on extracting the attribute features of the reference/target image and the modification text (described in Section 3.2.1). The second module targets composing the extracted attribute features of the reference image and the modification text, where the internal conflicting relationship between the multimodal query is modeled with the explicit guidance of the target-query relationship (detailed in Section 3.2.2). The third module aims to optimize the model with not only the conventional batch-based classification loss but also a newly proposed batch-based target similarity-guided matching degree regularization (explained in Section 3.2.3). We now detail each module of TG-CIR.


\subsubsection{\textbf{Attribute Feature Extraction.}}

As aforementioned, we aim to achieve the multimodal query composition through the ``\textit{keep-and-replace}'' paradigm. To fulfill this, we need to first obtain the unified attribute features of the reference image and the modification text, and then \textit{keep} the unchanged attributes of the reference image as well as \textit{replace} the conflicting attributes of the reference image with those of the modification text. Specifically, we adopt the contrastive language-image pre-training model (CLIP)~\cite{clip} as the attribute feature extraction backbone, which has shown remarkable success in the context of CIR~\cite{clip4cir}. Since it has been proven that both global and local features contribute to the CIR, we conduct both global and local attribute features extraction.
\textbf{Global Attribute Features Extraction.} In this part, we resort to the last-layer output vector of CLIP, which captures the overall feature of the image and the text, as the global feature basis and employ disentangled representation learning to derive the global attribute features of the three elements (\textit{i.e.},
reference image, modification text, and target image) involved in CIR. 
As the three elements share the same attribute feature extraction process, here we take that of the reference image as an example.
Let $\mathbf{f}_r \in \mathbb{R}^{D}$ denote the global feature of the reference image obtained by the last layer of the CLIP. Suppose there are $P$ global attributes. We then deploy $P$ learnable condition masks~\cite{zhengnamm} on $\mathbf{f}_r$ to derive the global attribute features of the reference image as follows,
\begin{equation}
    \mathbf{u}_{r}^{i} = \mathbf{f}_{r} \odot \mathbf{C}_{i},
    \label{eq2}
\end{equation}
where $\mathbf{C}_{i} \in \mathbb{R}^{D}$ denotes the $i$-th condition mask, and $\odot$ refers to the element-wise multiplication. $\mathbf{u}_{r}^{i}$ is the $i$-th global attribute feature of the reference image.


\textbf{Local Attribute Features Extraction.} In this part, we adopt the image/text tokens produced by the \mbox{second-to-last} layer of CLIP as the local feature basis for the image/text. For a given image, each token refers to a local region, while for a given text, each token corresponds to a word. Intuitively, each local attribute is correlated to only a few regions or words. Therefore, different from the global attribute features extraction, we propose to adaptively aggregate the attribute-related tokens to derive the local attribute features. Here we also take the process for the reference image as an example. 
Let $\mathbf{X}_r \in \mathbb{R}^{L \times D'}$ denote the local feature of the reference image obtained by the \mbox{second-to-last} layer of CLIP, where $L$ represents the number of output tokens for the reference image and $D'$ refers to the feature dimension. To facilitate the subsequent concatenation of the global and local attribute features, we first apply a fully connected layer $\operatorname{FC}_{I}\left( \cdot \right)$ to ensure the local feature dimension is the same as the global feature dimension as follows,
\begin{equation}
    \overline{\mathbf{X}}_{r} = \operatorname{FC}_{I} \left( {\mathbf{X}}_{r} \right),
    \label{eq3}
\end{equation}
where $\overline{\mathbf{X}}_{r} \in \mathbb{R}^{L \times D}$ is the projected local feature basis.
Then suppose there are $Q$ local attributes, and we accordingly design $Q$ feature aggregation functions, where the $j$-th feature aggregation function $\mathcal{A}_{j}$ is formulated as follows,
\begin{equation}
\left\{\begin{aligned}
\mathbf{s}_{r}^{j} &= \sigma \left( \operatorname{Conv}^{j}\left( \overline{\mathbf{X}}_{r} \right) \right), \\
\mathbf{v}_{r}^{j} &= {\mathbf{s}_{r}^{j}}^\top \otimes  \overline{\mathbf{X}}_{r} ,
\end{aligned}\right.\label{eq4}
\end{equation}
where $\mathbf{v}_{r}^{j} \in \mathbb{R}^{D}$ represents the $j$-th local attribute feature, $\operatorname{Conv}^{j}\left(\cdot\right)$ denotes the $1 \times 1$ convolution for the $j$-th feature aggregation, and $\sigma$ is the Sigmoid activation function. $\mathbf{s}_{r}^{j} \in \mathbb{R}^{1 \times L}$ is the weight vector obtained by the $1 \times 1$ convolution for aggregating the image tokens to derive the $j$-th local attribute feature. $\otimes$ is the matrix product, which acts as the weighted pooling over the image tokens.  

Finally, we cascade the global and local attribute features to derive the final attribute features of the reference image, denoted as $\mathbf{E}_{r} = \left[\mathbf{u}_{r}^{1}, \mathbf{u}_{r}^{2}, \cdots, \mathbf{u}_{r}^{P},  \mathbf{v}_{r}^{1}, \mathbf{v}_{r}^{2}, \cdots, \mathbf{v}_{r}^{Q}\right] \in \mathbb{R}^{K \times D}$, where $K=P+Q$. In the same way, we can obtain the attribute features of the modification text, denoted as $\mathbf{E}_{m}\in \mathbb{R}^{K \times D}$, and those of the target image, denoted as $\mathbf{E}_{t}\in \mathbb{R}^{K \times D}$. Note that to ensure the extracted attribute features of the three elements being aligned to facilitate the ``\textit{keep-and-replace}'' paradigm, we use the same $P$ condition masks and the same $Q$ feature aggregation functions for processing the three elements. 


\textit{Orthogonal regularization.} As there is no direct supervision signals (e.g., attribute labels) for the attribute feature learning,  we introduce the orthogonal regularization to guarantee that different attribute features can indeed represent different visual attributes and hence promote the attribute feature learning. Formally, the orthogonal regularization can be given as follows,
\begin{equation}
    \mathcal{L}_{ortho} =  \left\| \mathbf{E}_{r} {\mathbf{E}^{\top}_{r}}  - \mathbf{I} \right\|_{F}^{2} + \left\|\mathbf{E}_{m} {\mathbf{E}^{\top}_{m}}  - \mathbf{I} \right\|_{F}^{2} + \left\|\mathbf{E}_{t} {\mathbf{E}^{\top}_{t}}  - \mathbf{I} \right\|_{F}^{2},
    \label{eq12}
\end{equation}
where $\mathbf{I} \in \mathbb{R}^{K \times K}$ is the identity matrix, and $\left\| \cdot \right\|_{F}$ refers to the Frobenius norm of the matrix.

\subsubsection{\textbf{Target-Query Relationship-Guided Multimodal Query Composition.}}
As aforementioned, we aim to achieve an effective multimodal query composition by modeling the {conflicting relationship} between the multimodal query via the ``\textit{keep-and-replace}'' strategy. To fulfill this, intuitively, we can learn the \textit{keep mask} and \textit{replace mask}, which are responsible for \textit{keeping} the unchanged attributes of the reference image  and \textit{replacing} its conflicting attributes with those expressed by the modification text, respectively. 
As a major novelty, we propose to exploit the target-query relationship, whose benefit in promoting the {conflicting relationship} modeling has been shown in Figure~\ref{fig:intro}, to guide the {keep mask} and {replace mask} learning and hence improve the multimodal query composition effect.

Considering that the target image is only available in the training stage, we design this module with two multimodal query composition branches: (1) target-free student composition branch and (2) target-based teacher composition branch. The former student branch conducts the multimodal query composition only with the input multimodal query.  Differently, apart from the input multimodal query, the latter teacher branch additionally involves the target image for guiding the conflicting relationship modeling and hence promoting the multimodal query composition. In the training stage, the student branch will mimic the teacher branch in terms of {conflicting relationship} modeling and hence gain better multimodal query composition performance. Ultimately, the student branch can be directly used in the testing stage.


\textbf{{Target-free Student Composition.}} 
In this branch, we first feed the multimodal query into a multi-layer perception to learn the {keep mask} for multimodal query composition, and then based on that derive the {replace mask} as follows,
\begin{equation}
\left\{\begin{aligned}
\hat{\mathbf{m}}_{k} &= \sigma \left( \operatorname{MLP}_{s}\left( \left[ \mathbf{E}_{r}, \mathbf{E}_{m} \right] \right) \right), \\
\hat{\mathbf{m}}_{r} &= 1 - \hat{\mathbf{m}}_{k}, \\
\end{aligned}\right.\label{eq6}
\end{equation}
where $\sigma$ is the Sigmoid activation function. $\hat{\mathbf{m}}_{k} \in \mathbb{R}^{K}$ denotes the {keep mask} learned by the student branch, the $i$-th element of which refers to the probability that the $i$-th attribute feature of the reference image should be kept.
$\hat{\mathbf{m}}_{r}$ refers to the corresponding {replace mask} derived by ``$1-$\textit{keep mask}''. Intuitively, the higher the probability an attribute feature of the reference image should be kept, the lower the probability that the corresponding feature of the modification text should be considered for feature replacement.
Accordingly, the final composed query features are derived as follows,
\begin{equation}
    \hat{\mathbf{E}}_{c} = \hat{\mathbf{m}}_{k} \odot \mathbf{E}_{r} + \hat{\mathbf{m}}_{r} \odot \mathbf{E}_{m}. \label{eq7}
\end{equation}

\textbf{{Target-based Teacher Composition.}} 
As illustrated in Figure~\ref{fig:intro}, the {conflicting relationship} between the multimodal query can be modeled by referring to the relationship between the target image and the multimodal query. 
In particular, we propose to learn the {keep mask} according to the {consistency relationship} between the target image and the reference image, which intuitively reflects the unchanged attributes of the reference image. Meanwhile, to avoid the error accumulation, instead of obtaining the \textit{replace mask} through the ``$1-$\textit{keep mask}'' as the student branch,
we learn the {replace mask} directly according to the  {consistency relationship} between the target image and the modification text, which indicates the attributes of the reference image that need to be replaced. 
Formally, in this branch, we derive the composed features as follows,
\begin{equation}
\left\{\begin{aligned}
\tilde{\mathbf{m}}_{k} &= \sigma \left( \operatorname{MLP}_{t1}\left( \left[ \mathbf{E}_t, \mathbf{E}_r \right] \right) \right), \\
\tilde{\mathbf{m}}_{r} &= \sigma \left( \operatorname{MLP}_{t2}\left( \left[ \mathbf{E}_t, \mathbf{E}_m \right] \right) \right), \\
\tilde{\mathbf{E}}_{c} &= \tilde{\mathbf{m}}_{k} \odot \mathbf{E}_{r} + \tilde{\mathbf{m}}_{r} \odot \mathbf{E}_{m},
\end{aligned}\right.\label{eq8}
\end{equation}
where $\tilde{\mathbf{m}}_{k} \in \mathbb{R}^{K}$ and $\tilde{\mathbf{m}}_{r} \in \mathbb{R}^{K}$ represent the {keep mask} and the {replace mask} learned by the teacher branch, respectively.
Then, similar to the student branch, we adopt the following regularization to promote the two masks learning,
\begin{equation}
    \mathcal{L}_{mask}^{tea} = \left\| \tilde{\mathbf{m}}_{r}, \mathbf{1} - \tilde{\mathbf{m}}_{k} \right\|^{2},
\label{eq9}
\end{equation}
where $\left\| \cdot, \cdot \right\|^{2}$ denotes the mean square error function.

\textit{Composition Knowledge Distillation.} We then expect the student branch that can be applied in the testing stage to learn from the teacher branch that additionally involves the target image for promoting the multimodal query composition effect. As the key to our multimodal query composition lies in the two masks learning, we thus  
enforce the \textit{keep mask} and \textit{replace mask} learned by the student branch to be as similar as possible to those by the teacher branch. Formally, we have the following regularization for composition knowledge distillation,
\begin{equation}
    \mathcal{L}_{ckd} =  \left\|\tilde{\mathbf{m}}_{k} - \hat{\mathbf{m}}_{k}   \right\|^{2} + 
    \left\| \tilde{\mathbf{m}}_{r} - \hat{\mathbf{m}}_{r}  \right\|^{2}.
    \label{eq10}
\end{equation}

\subsubsection{\textbf{Target Similarity Distribution-Guided Metric Learning.}}
Having obtained the target image features and the composed query features, we can conduct the metric learning for CIR.
On one hand, we employ the commonly used batch-based classification loss as the fundamental optimization loss for metric learning. Essentially, the batch-based classification loss aims to enforce the composed query features close to the ground-truth target image features in the metric space, namely, maximizing the matching degree between the multimodal query and the target image. In fact, in our context, where the composed query and the target image are represented by multiple attribute features, we have two strategies for the batch-based classification loss design: 1) late fusion and 2) early fusion. The late fusion strategy first calculates the matching degree between the composed query and the target image on each attribute, and then sums the matching degrees on all attributes to derive the overall query-target matching degree. Differently, the early fusion strategy first fuses the multiple attribute features of the composed query and target image to derive their corresponding overall representation, and then based on that computing their matching degree. Intuitively, the late fusion strategy is more effective, while the early fusion strategy is more efficient. In our model, the teacher branch is used only for guiding the student branch during the training stage, while the student branch needs to be used in the testing stage. Therefore, we adopt the late fusion strategy for the teacher branch to promote the model effectiveness, while the early fusion strategy for the student branch to boost the model efficiency in the testing stage. Mathematically, we have the following two loss functions for optimizing the teacher branch and the student branch, respectively,
\begin{equation}
\begin{small}
\left\{\begin{aligned}
\mathcal{L}_{rank}^{tea} &= \frac{1}{B} \sum_{i=1}^{B} -\log \left\{ \frac{\exp \left\{ \left\{ \sum_{k=1}^{K} \operatorname{s}\left( \tilde{\mathbf{E}}_{ci}\left[k\right], \mathbf{E}_{ti}\left[k\right] \right) \right\}  / \tau\right\}}{ \sum_{j=1}^{B} \exp \left\{ \left\{ \sum_{k=1}^{K} \operatorname{s}\left( \tilde{\mathbf{E}}_{ci}\left[k\right], \mathbf{E}_{tj}\left[k\right] \right) \right\} / \tau \right\}        } \right\},\\
\mathcal{L}_{rank}^{stu} &= \frac{1}{B} \sum_{i=1}^{B} -\log \left\{ \frac{\exp \left\{ \operatorname{s} \left( \hat{\bm{\psi}}_{ci} , {\bm{\psi}}_{ti} \right)  / \tau\right\}}{ \sum_{j=1}^{B} \exp \left\{ \operatorname{s} \left( \hat{\bm{\psi}}_{ci}, {\bm{\psi}}_{tj} \right) / \tau \right\}  } \right\},
\end{aligned}\right.
    \label{eq12}
\end{small}
\end{equation}
where the subscript $i$ refers to the $i$-th triplet sample in the \mbox{mini-batch}, 
$B$ is the batch size, $\operatorname{s}(\cdot, \cdot)$ denotes the cosine similarity function, and $\tau$ is the temperature factor. 
$\tilde{\mathbf{E}}_{ci}\left[k\right]$ and ${\mathbf{E}}_{ti}\left[k\right]$ denote the $k$-th row of the two matrices, \textit{i.e.}, the $k$-th attribute feature.
$\hat{\bm{\psi}}_{ci}$ and ${\bm{\psi}}_{ti}$ stand for the composed feature vector of the multimodal query and the feature vector of the target image, respectively, which are derived by exerting the mean pooling operation to the composed feature matrix $\hat{\mathbf{E}}_{ci}$ and the target image feature matrix ${\mathbf{E}}_{ti}$ along the row.

On the other hand, we resort to the similarity distribution between the ground-truth target image feature and the batch candidates' features to regularize the metric learning of the student branch. Suppose that each batch has $B$ training triplets. We then compute the visual similarity distribution for each ground-truth target image with all the $B$ candidate target images in the same batch. Let $\mathbf{p}_i^{t}=\left[p_{i1}^{t},p_{i2}^{t},\cdots,p_{iB}^{t}\right]$ denote the batch-based visual similarity distribution regarding the $i$-th ground-truth target image, where $p_{ij}^{t}$ refers to the normalized visual similarity between the $i$-th ground-truth target image and the $j$-th candidate target image. Formally, we define $p_{ij}^{t}$ as follows,
\begin{equation}
p_{ij}^{t} = \frac{\exp \left\{ \left\{ \sum_{k=1}^{K} \operatorname{s}\left( {\mathbf{E}}_{ti}\left[k\right], \mathbf{E}_{tj}\left[k\right] \right) \right\}  / \tau\right\}}{ \sum_{b=1}^{B} \exp \left\{ \left\{ \sum_{k=1}^{K} \operatorname{s}\left( {\mathbf{E}}_{ti}\left[k\right], \mathbf{E}_{tb}\left[k\right] \right) \right\} / \tau \right\}        },
\label{eq13}
\end{equation}
Intuitively, the batch-based visual similarity distribution for each ground-truth target image can guide the matching degree learning for the corresponding triplet. Specifically, the more similar the candidate target image is to the ground-truth one, the higher the matching degree should be assigned to it for the given multimodal query. Accordingly, we also calculate the batch-based matching degree distribution for the multimodal query in each triplet. Let $\mathbf{p}_i^{c}=\left[p_{i1}^{c},p_{i2}^{c},\cdots,p_{iB}^{c}\right]$ denote the batch-based matching degree distribution regarding the $i$-th multimodal query, where $p_{ij}^{c}$ refers to the normalized matching degree between the $i$-th query and the $j$-th candidate target image. Following Eqn.~(\ref{eq12}), $p_{ij}^{c}$ can be obtained as follows,
\begin{equation}
p_{ij}^{c} = \frac{\exp \left\{ \operatorname{s} \left( \hat{\bm{\psi}}_{ci} , {\bm{\psi}}_{tj} \right)  / \tau\right\}}{ \sum_{b=1}^{B} \exp \left\{ \operatorname{s} \left( \hat{\bm{\psi}}_{ci}, {\bm{\psi}}_{tb} \right) / \tau \right\}  }.
\label{eq15}
\end{equation}
Then to guide the metric learning, we design the batch-based target similarity-guided matching degree regularization with the Kullback Leibler (KL) Divergence between $\mathbf{p}_i^{c}$ and $\mathbf{p}_i^{t}$ as follows,
\begin{equation}
\mathcal{L}_{kl}= \frac{1}{B}\sum_{i=1}^{B}D_{KL}\left({\mathbf{p}}_i^{t} \| {\mathbf{p}}_i^{c}\right)= \frac{1}{B}\sum_{i=1}^{B}\sum_{j=1}^{B} p_{ij}^{t} \log \frac{p_{ij}^{t}}{p_{ij}^{c}}.
\label{eq14}
\end{equation}

\textit{Optimization.} Integrating the three key modules, we parameterize the final objective function for our TG-CIR as follows,
\begin{small}
\begin{equation}
\mathbf{\Theta^{*}}=
\underset{\mathbf{\Theta}}{\arg \min } \left( {\mathcal{L}}_{rank}^{stu} + \lambda {\mathcal{L}}_{rank}^{tea} + \eta \mathcal{L}_{mask}^{tea} + \mu \mathcal{L}_{ortho} + \nu \mathcal{L}_{ckd} + \kappa \mathcal{L}_{kl} \right),\label{eq16}
\end{equation}
\end{small}where $\mathbf{\Theta}$ denotes the \mbox{to-be-learned} parameters in TG-CIR, $\lambda$, $\eta$, $\mu$, $\nu$, and $\kappa$ are the \mbox{trade-off} \mbox{hyper-parameters}. 

\section{Experiment}
In this section, we first introduce the experimental settings and then provide the experiment results and corresponding analyses.
 
\subsection{Experimental Settings}
\subsubsection{Datasets.}
We chose three public datasets for evaluation, including two fashion-domain datasets FashionIQ~\cite{fashioniq} and Shoes~\cite{shoes}, as well as an open-domain dataset CIRR~\cite{cirr}. 




\subsubsection{Implementation Details.}
TG-CIR utilizes pre-trained CLIP~\cite{clip} (ViT-B/$16$ version) as the backbone. We trained TG-CIR by AdamW optimizer with an initial learning rate of $1e-4$. Specifically, the learning rate of parameters in CLIP is set to $1e-5$ for better convergence. 
The learning rate decays by a factor of $0.1$ at the $5$-th and $10$-th epoch.
The dimensions of the global/local attribute features $D$ are set to $512$, and the batch size $B$ is set to $64$.
Regarding the number of attribute features, we set $Q$ to $8$, $6$, and $8$ for FashionIQ, Shoes, and CIRR through hyperparameter tuning, respectively.
We empirically set $P = \frac{Q}{2}$ to facilitate parameter setting, as the number of global attributes is usually fewer than that of local attributes. The temperature factor $\tau$ in Eqn.~(\ref{eq12}), Eqn.~(\ref{eq13}), and Eqn.~(\ref{eq15}) is set to $0.1$ for both FashionIQ and Shoes, and $0.05$ for CIRR according to the grid search among $\left[0.01, 0.05, 0.1, 0.5\right]$.
Regarding the trade-off hyper-parameters, $\lambda$ and $\eta$ are empirically set to $1.0$, while $\mu$, $\nu$, and $\kappa$ are set through grid search among $\left[ 0.05, 0.1 \right]$, $\left[ 1, 5, 10, 50 \right]$, and $\left[0.01, 0.05, 0.1, 0.5\right]$, respectively. Finally, we set $\mu = 0.1$, $\nu = 10$, and $\kappa = 0.5$ for FashionIQ, $\mu = 0.05$, $\nu = 5$, and $\kappa = 0.5$ for Shoes, and $\mu = 0.1$, $\nu = 1$, and $\kappa = 0.1$ for CIRR, respectively.
\subsubsection{Evaluation.} We used the standard evaluation protocol for each dataset and reported the Recall@$k$ (R@$K$ for short) metric for a fair comparison. 
For FashionIQ, we followed the evaluation metric used in the companion challenge~\cite{fashioniq}, namely, R@$10$ and R@$50$ for each of the three categories.
For Shoes, we reported R@$1$, R@$10$, R@$50$, and their average as the same as previous efforts~\cite{crr,clvcnet,amc}.
For CIRR, following~\cite{cirr,artemis}, we reported R@$k$ ($k=1,5,10,50$), R$_{subset}$@$k$ ($k=1,2,3$), and the average of R@$5$ and R$_{subset}$@$1$. 

\begin{table*}[t!]
    \centering
    \caption{Performance comparison on FashionIQ and Shoes with respect to R@$k$($\%$). The best results are in boldface, while the second-best results are underlined. The missing results of some methods are because they did not report their results on the Shoes dataset. The last row indicates the performance improvements by TG-CIR over the best baseline.
    }
    \vspace{-0.7em}
    \begin{tabular}{l|cc|cc|cc|cc|c|c|c|c}
    \hline 
    \multirow{3}{*}{Method} &\multicolumn{8}{c|}{FashionIQ} &\multicolumn{4}{c}{Shoes} \\ \cline{2-13} 
    & \multicolumn{2}{c|}{Dresses} & \multicolumn{2}{c|}{Shirts} & \multicolumn{2}{c|} {Tops\&Tees} & \multicolumn{2}{c|}{Avg} & \multirow{2}{*}{R@$1$}  & \multirow{2}{*}{R@$10$} &\multirow{2}{*}{R@$50$} & \multirow{2}{*}{Avg} \\ \cline{2-9}  & R@$10$ & R@$50$ & R@$10$ & R@$50$ & R@$10$ & R@$50$ & R@$10$ & R@$50$ & & & \\
    \hline \hline 
    TIRG~\cite{tirg} \small{\textcolor{gray}{(CVPR'19)}} & $14.87$ & $34.66$ & $18.26$ & $37.89$ & $19.08$ & $39.62$ & $17.40$ & $37.39$ &$12.60$ & $45.45$ & $69.39$ &$42.48$\\
    VAL~\cite{val} \small{\textcolor{gray}{(CVPR'20)}} & $21.12$ & $42.19$ & $21.03$ & $43.44$ & $25.64$ & $49.49$ & $22.60$ & $45.04$ & $16.49$ & $49.12$ & $73.53$ & $46.38$\\
    CIRPLANT~\cite{cirr} \small{\textcolor{gray}{(ICCV'21)}} & $17.45$ & $40.41$ & $17.53$ & $38.81$ & $21.64$ & $45.38$ & $18.87$ & $41.53$ & $-$ & $-$ & $-$ & $-$\\
    CosMo~\cite{cosmo} \small{\textcolor{gray}{(CVPR'21)}}  & $25.64$ & $50.30$ & $24.90$ & $49.18$ & $29.21$ & $57.46$ & $26.58$ & $52.31$ &$16.72$ & $48.36$ & $75.64$ &$46.91$\\
    DATIR~\cite{datir} \small{\textcolor{gray}{(ACM MM'21)}}  & $21.90$ & $43.80$ & $21.90$ & $43.70$ & $27.20$ & $51.60$ & $23.70$ & $46.40$ &$17.20$ & $51.10$ & $75.60$ &$47.97$\\
    MCR~\cite{hffca} \small{\textcolor{gray}{(ACM MM'21)}} & $26.20$ & $51.20$ & $22.40$ & $46.00$ & $29.70$ & $56.40$ & $26.10$ & $51.20$ &$17.85$ & $50.95$ & $77.24$ &$48.68$\\
    CLVC-Net~\cite{clvcnet} \small{\textcolor{gray}{(SIGIR'21)}} & $29.85$ & $56.47$ &$ 28.75$ & $54.76$ & $33.50$ & $64.00 $& $30.70$ &$ 58.41 $ &$17.64$ & $54.39$ & $79.47$ &$50.50$\\
    ARTEMIS~\cite{artemis} \small{\textcolor{gray}{(ICLR'22)}}  & $27.16$ & $52.40$ & $21.78$ & $43.64$ &$ 29.20 $& $54.83 $& $26.05 $& $50.29 $ &$18.72$ & $53.11$ & $79.31$ &$50.38$\\
    EER~\cite{tip22} \small{\textcolor{gray}{(TIP'22)}} & $30.02$ & $55.44$ & $25.32$ & $49.87$ & $33.20$ & $60.34$ & $29.51$ & $55.22$ & \underline{$20.05$} & $56.02$ & \underline{$79.94$} & $52.00$\\
    FashionVLP~\cite{fashionvlp} \small{\textcolor{gray}{(CVPR'22)}} & $32.42$ & $60.29$ & $31.89$ & $58.44$ & $38.51$ & $68.79$ & $34.27$ & $62.51$ & $-$ & $49.08$ & $77.32$ & $-$\\
    CRR~\cite{crr} \small{\textcolor{gray}{(ACM MM'22)}} & $30.41$ & $57.11$ & $30.73$ & $58.02$ & $33.67$ & $64.48$ & $31.60$ & $59.87$ & $18.41$ & $56.38$ & $79.92$ & $51.57$\\
    AMC~\cite{amc} \small{\textcolor{gray}{(TOMM'23)}} & $31.73$ & $59.25$ & $30.67$ & $59.08$ & $36.21$ & $66.60$ & $32.87$ & $61.64$ & $19.99$ & \underline{$56.89$} & $79.27$ & \underline{$52.05$}\\
    \hline 
    Clip4cir~\cite{clip4cir} \small{\textcolor{gray}{(CVPRW'22)}} & $33.81$ & $59.40$ & $39.99$ & $60.45$ & $41.41$ & $65.37$ & $38.32$ & $61.74$ & $-$ & $-$ & $-$ & $-$\\
    FAME-ViL\cite{fame_vil} \small{\textcolor{gray}{(CVPR'23)}} & \underline{$42.19$} & \underline{$67.38$} &\underline{$47.64$}& \underline{$68.79$} & \underline{$50.69$} & \underline{$73.07$} & \underline{$46.84$} &\underline{$ 69.75 $} & $-$ & $-$ & $-$ & $-$\\
    \hline \hline
    \textbf{TG-CIR} & $\mathbf{45.22}$ & $\mathbf{69.66}$ & $\mathbf{52.60}$ & $\mathbf{72.52}$ & $\mathbf{56.14}$ & $\mathbf{77.10}$ & $\mathbf{51.32}$ & $\mathbf{73.09}$ & $\mathbf{25.89}$ & $\mathbf{63.20}$ & $\mathbf{85.07}$ & $\mathbf{58.05}$\\
     Improvement(\%) & $\uparrow 7.18 $ & $\uparrow 3.38 $ & $\uparrow 10.41 $ & $\uparrow 5.42 $ & $\uparrow 10.75 $ & $\uparrow 5.52 $ & $\uparrow 9.56 $ & $\uparrow 4.79 $ & $\uparrow 29.13$ & $\uparrow 11.09$ & $\uparrow 6.42$ &$\uparrow 11.53$\\
    \hline
    \end{tabular}
    \label{tab:exp_fashioniq_shoes}
    \vspace{-0.5em}
\end{table*}

\begin{table*}[t!]
    \centering \caption{Performance comparison on CIRR with respect to R@$k$($\%$) and R$_{subset}$@$k$($\%$). The best results are in boldface, while the second-best results are underlined. The last row indicates the performance improvements by TG-CIR over the best baseline.}
    \vspace{-0.7em}
    \begin{tabular}{l|c|c|c|c|c|c|c|c}
    \hline 
    \multirow{2}{*}{Method} &\multicolumn{4}{c|}{\textbf{R@$k$}} &\multicolumn{3}{c|}{\textbf{R$_{subset}$@$k$}} & \multirow{2}{*}{Avg} \\ \cline{2-8}
    & $k=1$ & $k=5$ & $k=10$ & $k=50$ & $k=1$ & $k=2$ & $k=3$ \\
    \hline \hline 
    TIRG~\cite{tirg} \small{\textcolor{gray}{(CVPR'19)}} & $14.61$ & $48.37$ & $64.08$ & $90.03$ & $22.67$ & $44.97$ & $65.14$ & $35.52$\\
    ARTEMIS~\cite{artemis} \small{\textcolor{gray}{(ICLR'22)}} & $16.96$ & $46.10$ & $61.31$ & $87.73$ &$ 39.99 $& $62.20 $& $75.67 $ & $43.05$\\
    CIRPLANT~\cite{cirr} \small{\textcolor{gray}{(ICCV'21)}} & $15.18$ & $43.36$ & $60.48$ & $87.64$ & $33.81$ & $56.99$ & $75.40$ & $38.59$\\
    \hline
    Clip4cir~\cite{clip4cir} \small{\textcolor{gray}{(CVPRW'22)}} & \underline{$38.53$} & \underline{$69.98$} &\underline{$81.86$}& \underline{$95.93$} & \underline{$68.19$} & \underline{$85.64$} & \underline{$94.17$} & \underline{$69.09$}\\
    \hline \hline
    \textbf{TG-CIR} & $\mathbf{45.25}$ & $\mathbf{78.29}$ & $\mathbf{87.16}$ & $\mathbf{97.30}$ & $\mathbf{72.84}$ & $\mathbf{89.25}$ & $\mathbf{95.13}$ & $\mathbf{75.57}$ \\
    Improvement(\%) & $\uparrow 17.44 $ & $\uparrow 11.87 $ & $\uparrow 6.47 $ & $\uparrow 1.43 $ & $\uparrow 6.82 $ & $ \uparrow 4.22 $ & $ \uparrow 1.02 $ & $ \uparrow 9.38$ \\ 
    \hline
    \end{tabular}
    \label{tab:exp_cirr}
    \vspace{-0.7em}
\end{table*}

\subsection{Performance Comparison}
We compared TG-CIR with the following baselines: TIRG\cite{tirg}, VAL\cite{val}, CIRPLANT\cite{cirr}, CosMo\cite{cosmo}, 
DATIR\cite{datir}, 
MCR\cite{hffca}, CLVC-Net\cite{clvcnet}, ARTEMIS\cite{artemis}, EER\cite{tip22}, FashionViL\cite{fashionvil}, CRR\cite{crr}, AMC\cite{amc},  Clip4cir\cite{clip4cir}, and FAME-ViL\cite{fashionvil}. Notably, the former twelve baselines use traditional models like ResNet~\cite{resnet} and LSTM~\cite{lstm} as the feature extraction backbone, while the latter two take advantage of the multimodal pre-trained large model CLIP.  
Tables~\ref{tab:exp_fashioniq_shoes} and ~\ref{tab:exp_cirr} summarize the performance comparison on the three datasets. It is worth mentioning that since the CIRR dataset is newly released, only a few baselines' results are available.  Our observations are presented in the following. 
1) TG-CIR consistently outperforms all baselines over the three datasets from both fashion domain and open domain by a significant margin. 
Specifically, TG-CIR achieves $9.56\%$ improvements over the best baseline for R@$10$ on FashionIQ-Avg, $29.13\%$ for R@$1$ on Shoes, and $17.44\%$ for R@$1$ on CIRR.
These results demonstrate the generalization ability and versatility of our model. 
2) Methods using traditional feature extraction backbones (top of the tables) generally perform worse than those using CLIP (bottom of the tables). This suggests the benefit of using the pretrained CLIP model for image/text feature extraction in CIR.
3) FAME-ViL achieves the second-best performance on FashionIQ. This may be due to the fact that this model is optimized through a multi-task learning approach on multiple datasets within the fashion domain.
In contrast, our model is not limited to a specific domain and can achieve promising CIR performance for various domains.

\subsection{Ablation Study}
To verify the importance of each module in our model, we compared TG-CIR with its following derivatives.
\begin{itemize}[leftmargin=8.5pt]
    \item \textbf{Local-AttriFea\_Only} and \textbf{Global-AttriFea\_Only:} To investigate the necessity of extracting the attribute features from both local and global perspectives, we employed only the local or global attribute features in $\mathbf{E}_{r}$, $\mathbf{E}_{m}$, and $\mathbf{E}_{t}$, respectively.
    \item \textbf{w/o\_ortho:} To verify the effect of orthogonal regularization, we removed $\mathcal{L}_{ortho}$ in Eqn.~(\ref{eq16}) by setting $\mu=0$.
    \item \textbf{w/o\_target\_guide:} To check the effect of leveraging the target image for guiding the multimodal query composition and metric learning, we only utilized the target-free student composition in the multimodal query composition and the batch-based classification loss in the metric learning by setting $\lambda$, $\eta$, $\nu$, and $\kappa$ as $0$ in Eqn.~(\ref{eq16}).
    \item \textbf{w/o\_target\_guide\_c:} To further investigate the effect of the target-based teacher composition, we removed the guidance for learning the {keep mask} and replace mask from the teacher branch to the student branch by setting $\nu=0$ in Eqn.~(\ref{eq16}).
    \item \textbf{w/o\_target\_guide\_m:} To explore the effect of the target similarity-guided matching degree regularization, we removed $\mathcal{L}_{kl}$ from Eqn.~(\ref{eq16}) by setting $\kappa=0$.
\end{itemize}

Table~\ref{exp:ablation} shows the ablation results of TG-CIR on three datasets. From this table, we gained the following observations. 
1) Both Local-AttriFea\_Only and Global-AttriFea\_Only are inferior to TG-CIR, which validates the importance of incorporating both the  local and global attribute features in the context of CIR.
2) TG-CIR surpasses w/o\_ortho, indicating that the designed orthogonal regularization is indeed helpful to guarantee the attribute features
independence and promote the attribute feature extraction.
3) w/o\_target\_guide performs worse than TG-CIR, which validates the effectiveness of exploiting the visual information of the target image in the CIR task.
4) w/o\_target\_guide\_c is inferior to TG-CIR, which suggests that it is essential to utilize the target image to promote the conflicting relationship modeling between the reference image and the modification text in the multimodal query composition.
5) TG-CIR outperforms w/o\_target\_guide\_m, which confirms the benefit of incorporating the visual similarity distribution to regularize the metric learning process.

\begin{table}[t!]
    \centering \caption{Ablation study on FashionIQ, Shoes, and CIRR. 
    }
    \vspace{-0.3em}
    \begin{tabular}{l|c|c|c|c}
    \hline 
    \multirow{2}{*}{Method} &\multicolumn{2}{c|}{FashionIQ-Avg} & Shoes & CIRR \\ \cline{2-5}
    & R@$10$ & R@$50$ & Avg  & Avg  \\
    \hline \hline 
    Local-AttriFea\_Only& $ 41.92 $ & $ 67.37 $ & $ 52.35 $ & $ 55.68$ \\
    Global-AttriFea\_Only & $ 49.50 $ & $ 72.89 $ & $ 56.31 $ & $ 74.50$ \\
    w/o\_ortho & $50.77$ & $72.23$ & $ 57.50 $ &$ 75.04$ \\
    w/o\_target\_guide & $48.84$ & $72.28$ & $ 55.80 $ & $ 72.62$ \\
    w/o\_target\_guide\_c & $50.00$ & $72.82$ & $55.84 $ & $ 74.17 $ \\
    w/o\_target\_guide\_m & $48.95$ & $72.31$ & $ 56.33 $ & $ 74.58$ \\
    \textbf{TG-CIR} & $\mathbf{51.32}$ & $\mathbf{73.09}$ & $\mathbf{58.05}$ & $\mathbf{75.57}$  \\
    \hline
    \end{tabular}
    \vspace{-0.3em}
    \label{exp:ablation}
\end{table}

\begin{figure}[h]
    \centering
	\includegraphics[width=1.0\linewidth]{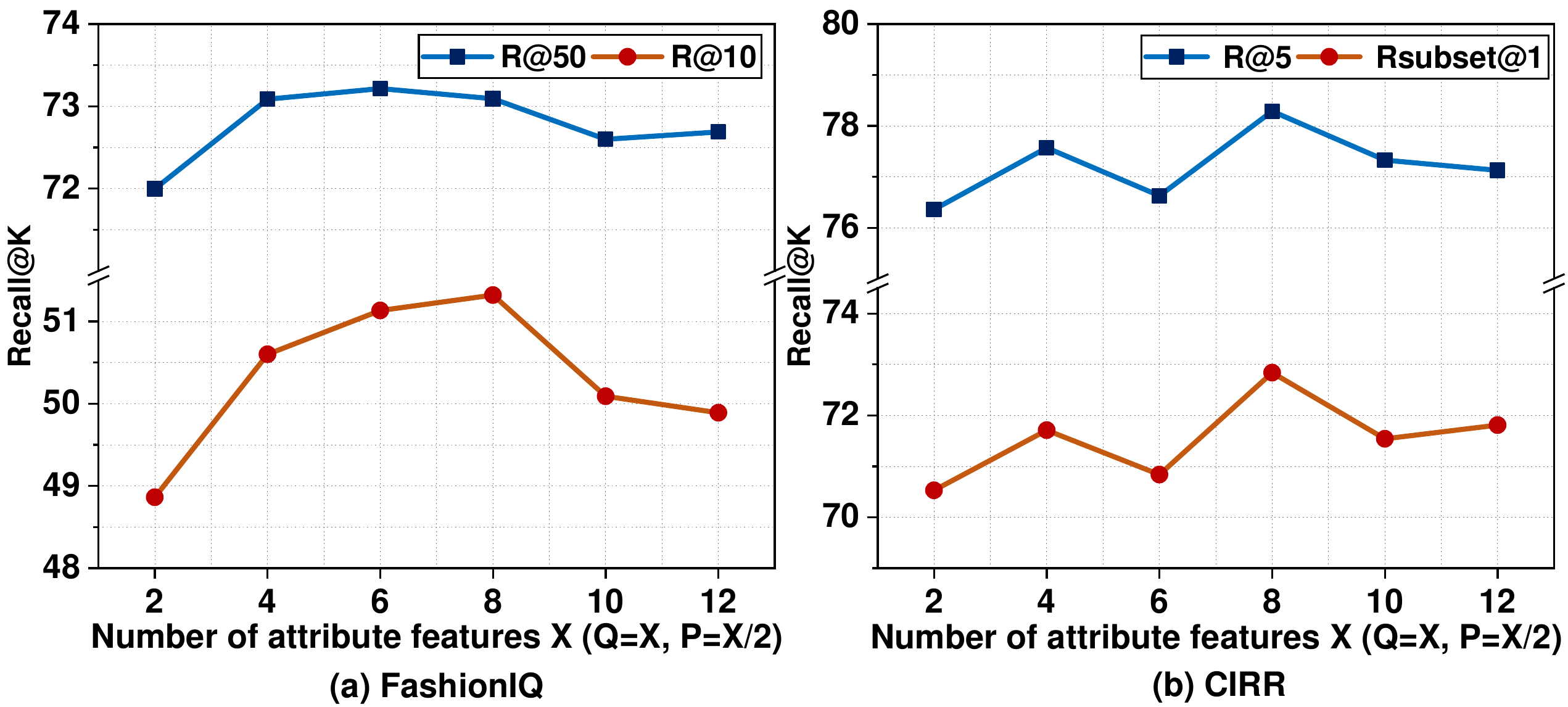}
    \vspace{-0.3em}
	\caption{Influence of the number of the attribute features on (a) FashionIQ and (b) CIRR.}
    \vspace{-0.5em}
	\label{fig:num_feature}
\end{figure}

To provide more insight into the effect of attribute features, we illustrate the performance of TG-CIR with respect to different numbers of attribute features in Figure~\ref{fig:num_feature}. Due to the limited space, we show the results on FashionIQ and CIRR. Note that we set $P=\frac{Q}{2}$ to facilitate the parameter tuning. As can be seen, the performance of our model generally boosts with the increasing number of attribute features, followed by a decline. This is reasonable, as a greater number of attribute features can potentially cover a broader range of modification demands and enhance the CIR performance. However, since in practice the number of useful attributes is limited, too many attribute features would introduce redundancy and hence hurt the retrieval performance.

\begin{figure}[ht!]
    \centering
	\includegraphics[width=0.98\linewidth]{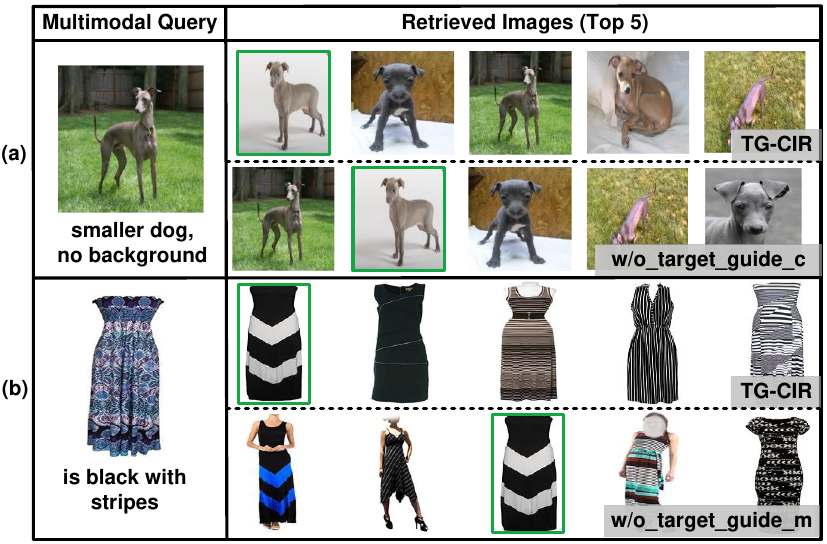}
    \vspace{-0.5em}
	\caption{Case study on (a) CIRR and (b) FashionIQ datasets.}
    \vspace{-1em}
	\label{fig:case_study}
\end{figure}
\subsection{Case Study}
Figure~\ref{fig:case_study} illustrates two \mbox{CIR} examples obtained by TG-CIR and two key derivatives
on the open-domain CIRR and fashion-domain FashionIQ. The top $5$ retrieved images are listed, where the green boxes indicate the target images. As shown in Figure~\ref{fig:case_study}(a), TG-CIR ranks the target image in the first place, while w/o\_target\_guide\_c ranks it in the second place. This indicates the importance of introducing the target-query relationship to guide the multimodal query composition. 
Regarding the case in Figure~\ref{fig:case_study}(b), TG-CIR successfully ranks the target image in the first place, while w/o\_target\_guide\_m fails. 
Meanwhile, we noticed that the top $5$ images retrieved by TG-CIR better meet the multimodal query (\textit{i.e.}, only changing the color and pattern of the reference image) than those retrieved by w/o\_target\_guide\_m. 
This suggests the benefit of utilizing the target visual similarity distribution to guide the metric learning process.



\section{Conclusion}

In this work, we present a novel target-guided composed image retrieval network to address the challenging CIR task, which exploits the target image to guide both multimodal query composition and metric learning for CIR. In particular, we first extracted the attribute features of reference/target image and modification text to facilitate the following ``\textit{keep-and-replace}'' paradigm based multimodal query composition, where a target-based teacher composition branch is employed to guide the learning of a target-free student composition branch. Moreover, we utilized the visual similarities between the ground-truth target image and candidate target images to guide the matching degrees learning between the multimodal query and candidate target images, thus boosting the metric learning effect. Extensive experiments have been conducted on three public datasets, and the results demonstrate the effectiveness of TG-CIR. In the future, we plan to extend our method to solve the multi-turn interactive image retrieval task, which is essential in multimodal dialogue systems and intelligent customer service robots.

\section*{Acknowledgments}
This work is supported by the Shandong Provincial Natural Science Foundation (No.:ZR2022YQ59). It is also in part based on research sponsored by Defense Advanced Research Projects Agency (DARPA) under agreement number HR0011-22-2-0047.

\clearpage
\balance

\bibliographystyle{ACM-Reference-Format}
\bibliography{reference}

\end{sloppypar}
\end{document}